\documentclass[conference,Letter]{IEEEtran}

\usepackage[utf8]{inputenc} 
\usepackage[T1]{fontenc}
\usepackage{url}
\usepackage{ifthen}
\usepackage{cite}
\usepackage[cmex10]{amsmath} 
\usepackage{amssymb}
\usepackage{graphicx}
\usepackage{subfigure}
\usepackage{epstopdf}

\newtheorem{thm}{Theorem}[section]

\newtheorem{exmp}[thm]{Example}

\begin{document}
\title{On the Optimality of Ali-Niesen Decentralized Coded Caching Scheme With and Without Error Correction} 


\author{%
  \IEEEauthorblockN{Nujoom Sageer Karat and B. Sundar Rajan}
  \IEEEauthorblockA{Department of Electrical Communication Engineering, Indian Institute of Science, Bengaluru 560012, KA, India \\
E-mail: \{nujoom,bsrajan\}@iisc.ac.in}
}

\maketitle

\begin{abstract}
 The decentralized coded caching was introduced in [M.~A. Maddah-Ali and U.~Niesen, ``Decentralized coded caching attains order-optimal memory-rate tradeoff,'' \emph{IEEE/ACM Trans. Networking}, vol.~23, no.~4, pp. 1029--1040, Aug. 2015] in which no coordination is required for the content placement. In this paper, with $N$ files and $K$ users, first it is shown that the decentralized coded caching scheme of  Ali and Niesen is optimal for $N \geq K$ using techniques from index coding. Next the case when the shared link is error prone is considered. For this case, an optimal error correcting delivery scheme is proposed for coded caching problems with decentralized placement. Expression for the worst case rate is obtained when there are finite number of transmissions in error. 
\end{abstract}
\section{INTRODUCTION}
Coded caching has gained interest recently as it is used to reduce the network traffic during peak hours \cite{MaN}. A coded caching scheme consists of two phases: a placement phase (prefetching phase) and a delivery phase. The local cache memory of each user is filled using the entire file database during the placement or prefetching phase. Delivery phase is carried out once the demands are revealed by the users. During placement phase some parts of files have to be judiciously cached at each user in such a way that the rate of transmission is reduced during the delivery phase.

 The fundamental scheme introduced in \cite{MaN} is a centralized coded caching scheme in which the placement is centrally coordinated, which limits its applicability. In real life scenarios, the identity or even the number of active users may not be known several hours in advance during the placement phase. There can be even change of networks from placement phase to delivery phase. Thus coordination in the placement phase may not be possible in some application scenarios. Decentralized coded caching scheme was introduced in \cite{MaN2}, which assumes no coordination during ithe placement phase. Using the scheme in \cite{MaN2}, coded-multicasting opportunities are still created in the delivery phase. Decentralized coded caching is widely applicable in other contexts like online coded caching \cite{PMN} and hierarchical coded caching \cite{KNMD}.

 The schemes in \cite{MaN, MaN2} use uncoded prefetching, i.e., each user stores a subset of the bits of the original
files. If coding is done during prefetching phase, then the prefetching is referred to as coded prefetching \cite{CFL,JV}. In \cite{YMA}, the scheme in \cite{MaN} is proven to be optimal under the restriction of uncoded placement. In this work, we prove the optimality of the delivery scheme in \cite{MaN2} when uncoded decentralized placement is used.

The case when the shared link is error prone is considered in \cite{KTR, KTR_2}. In this set up, the delivery phase is assumed to be error-prone and placement is assumed to be error-free. A similar model in which the delivery phase takes place over a packet erasure broadcast channel was considered in \cite{BWT}. In our work, we extend this to the case of decentralized caching scheme in \cite{MaN2}. Error correction at receivers can be achieved by increasing the number of transmissions. This paper addresses the problem of finding optimal linear error correcting delivery schemes which use minimum number of transmissions. By optimal error correcting delivery scheme, we mean a delivery scheme with minimum number of transmissions for a given cache placement scheme.
The main contributions of this paper are as follows.
\begin{itemize}
\item Optimality of the delivery scheme in \cite{MaN2} is proved for decentralized placement  for the number of files not less than the number of users using results from index coding (Section \ref{sec:optimality}).

\item Optimal error correcting delivery scheme is found for coded caching problems with decentralized placement \cite{MaN2} (Section \ref{sec:error_cor}).

\item Closed form expression for worst case rate is found for decentralized delivery scheme in the presence of finite number of transmission errors (Section \ref{sec:error_cor}). 
\end{itemize}

In this paper $\mathbb{F}_{q}$ denotes the finite field with $q$ elements, where $q$ is a power of a prime, and $\mathbb{F}^{*}_{q}$ denotes the  set of all nonzero elements of $\mathbb{F}_q$. The notation $[K]$ is used for the set $\{1,2,...,K\}$ for any integer $K$. For a $K \times N$ matrix $L$, $L_i$ denotes its $i$th row. For a set $S \subseteq [K]$, $L_S$ denotes the $ |S| \times N$ matrix obtained from $L$ by deleting the rows of $L$ which are not indexed by the elements of $S$. We denote $\bf{e_\textit{i}}$ $= (\underbrace{0,...,0}_{i-1},1,\underbrace{0,...,0}_{K-i}) \in \mathbb{F}^{n}_{q}$ as the unit vector having a one at the $i$th position and zeros elsewhere.

A linear $[n,k,d]_q$ code $\mathcal{C}$ over $\mathbb{F}_q$ is a $k$-dimensional subspace of $\mathbb{F}^{n}_{q}$ with minimum Hamming distance $d$. The vectors in $\mathcal{C}$ are called codewords. A matrix ${G}$ of size $k \times n$ whose rows are linearly independent codewords of $\mathcal{C}$ is called a generator matrix of $\mathcal{C}$. A linear $[n,k,d]_q$ code $\mathcal{C}$ can thus be represented using its generator matrix ${G}$ as,
$ \mathcal{C} = \{ {\bf{y}}G: {\bf{y}} \in \mathbb{F}^{k}_{q} \}.$ 
Let $N_{q}[k,d]$ denote the length of the shortest linear code over $\mathbb{F}_q$ which has dimension $k$ and minimum distance $d$.

\section{Preliminaries and Background}
In this section we review the basic results from error correcting index coding \cite{DSC} which are lates used in this paper to show the optimality of the decentralized scheme in \cite{MaN2} and to obtain an optimal  error correction scheme for the same. We also revisit the decentralized scheme \cite{MaN2} in brief and  error correcting coded caching terminologies from \cite{KTR}.

\subsection{Index Coding Problem}
The index coding problem with side information was introduced in \cite{BiK}. A single source has $n$ messages $x_1,x_2 \ldots , x_n$ where $x_i \in \mathbb{F}_{q}, ~\forall i \in [n].$ There are $K$ receivers, $R_1, R_2, \ldots, R_K$. Each receiver possesses a subset of messages as side information. Let $\mathcal{X}_i$ denote the set of  indices of the messages belonging to the side information of receiver $R_i$. The map $f:[K] \rightarrow [n] $ assigns receivers to indices of messages demanded by them. Receiver $R_i$ demands the messages $x_{f(i)}$, $f(i) \notin \mathcal{X}_i$ \cite{DSC}. The source knows the side information available to each receiver and has to satisfy the demand of each receiver in minimum number of transmissions. An instance of index coding problem can be completely characterized by a side information hypergraph \cite{AHLSW}. Given an instance of the index coding problem, finding the best \textit{scalar linear} binary index code is equivalent to finding the \textit{min-rank} of the side information hypergraph \cite{DSC}, which is known to be an NP-hard problem in general \cite{BBJ, RP, DSC2}. 

%

An index coding problem with $K$ receivers and $n$ messages can be represented by a hypergraph $\mathcal{H} (V,E)$, where $V=[n]$ is the set of vertices and $E$ is the set of hyperedges \cite{AHLSW}. Vertex $i$ represents the message $x_i$ and each hyperedge represents a receiver. In \cite{DSC}, the min-rank of a hypergraph $\mathcal{H}$ corresponding to index coding problem $\mathcal{I}$ over $\mathbb{F}_q$ is defined as,
\begin{equation*}
\kappa(\mathcal{I}) \triangleq
\min\{\text{rank}_q(\{{\mathbf{v_i}}
+{\mathbf{e_{f(i)}}}\}_{i\in [K]}): \\
{\mathbf{v_i}} \in \mathbb{F}^{n}_q, {\mathbf{v_i}} \triangleleft \mathcal{X}_i\},
\end{equation*}
where $\bf{v_i}$ $\triangleleft$ $\mathcal{X}_i$ denotes that $\bf{v_i}$ is the subset of the  support of $\mathcal{X}_i$; the support of a vector $\bf{u}$ $\in \mathbb{F}^{n}_{q}$ is defined to be the set $\{i\in [n]: u_i \neq 0  \}$. This min-rank defined above is the smallest length of scalar linear index code for the problem. A linear index code of length $N$ can be expressed as $XL$, where $L$ is an $n \times N$ matrix and $X = [x_1~x_2 \ldots ~x_n]$. The matrix $L$ is said to be the \textit{matrix corresponding to the index code}. 

For an undirected graph $\mathcal{G}$ = $(\mathcal{V},\mathcal{E})$, a subset of vertices $\mathcal{S}$ $\subseteq$ $\mathcal{V}$ is called an independent set if $\forall u, v \in \mathcal{S}$, $\{u,v\}$ $\notin$ $\mathcal{E}$. The size of a largest independent set in the graph $\mathcal{G}$ is called the independence number of $\mathcal{G}$. Dau {\it{et al}}. in \cite{DSC} extended the notion of independence number to the case of directed hypergraph corresponding to an index coding problem. For each receiver $R_i$, define the sets $$
\mathcal{Y}_i \triangleq [n] \setminus \bigg( \{f(i) \} \cup \mathcal{X}_i \bigg) $$
and
$$\mathcal{J(\mathcal{I})} \triangleq \cup_{i\in [K]} \{\{f(i)\} \cup Y_{i} : Y_i \subseteq \mathcal{Y}_i\}.$$
A subset $H$ of $[n]$ is called a generalized independent set in $\mathcal{H}$, if every nonempty subset of $H$ belongs to $\mathcal{J(\mathcal{I})}$. The size of the largest independent set in $\mathcal{H}$ is called the generalized independence number and is denoted by $\alpha (\mathcal{I})$. It is proved in \cite{KTR} that for any index coding problem,
\begin{equation}
\label{eq:alphaleqkappa}
\alpha (\mathcal{I}) \leq \kappa (\mathcal{I}).
\end{equation}

The quantities  $\alpha (\mathcal{I})$ and $\kappa(\mathcal{I})$ decide the bounds on the optimal length of error correcting index codes. The error correcting index coding problem  with side information was defined in \cite{DSC}. An index code is said to correct $\delta$ errors if after receiving at most $\delta$ transmissions in error, each receiver is able to decode its demand. A $\delta$-error correcting index code is represented as $(\delta, \mathcal{I})$-ECIC. An optimal linear $(\delta, \mathcal{I})$-ECIC over $\mathbb{F}_q$ is a linear $(\delta, \mathcal{I})$-ECIC over $\mathbb{F}_q$ of the smallest possible length $\mathcal{N}_{q}[\mathcal{I},\delta]$. 
Lower and upper bounds on  $\mathcal{N}_{q}[\mathcal{I},\delta]$ were established in \cite{DSC}. The lower bound is known as the $\alpha$-bound and the upper bound is known as the $\kappa$-bound.  
The length of an optimal linear $(\delta,\mathcal{I})$-ECIC over $\mathbb{F}_q$ satisfies
\begin{equation}
\underbrace{N_q[\alpha(\mathcal{I}), 2\delta + 1]~ \leq ~}_{\alpha\text{-bound}}  \mathcal{N}_{q}[\mathcal{I},\delta] \underbrace{~\leq~ N_q[\kappa(\mathcal{I}), 2\delta + 1]}_{\kappa\text{-bound}}. \label{eq:bds}
\end{equation}
The $\kappa$-bound is achieved by concatenating an optimal linear classical error correcting code and an optimal linear index code. Thus for any index coding problem, if $\alpha (\mathcal{I})$ is same as $\kappa(\mathcal{I})$, then concatenation scheme would give optimal error correcting index codes \cite{SaR, SagR, SSR, KSR}.

\subsection{Error Correcting Coded Caching Scheme}
Error correcting coded caching scheme was proposed in \cite{KTR}. The server is connected to $K$ users through a shared link which is error prone. 
The server has access to $N$ files $X_1, X_2, \ldots, X_N$, each of size $F$ bits. Every user has an isolated cache with memory $MF$ bits, where $M \in [0,N]$.  A prefetching scheme is denoted by ${\mathcal{M}}$. During the delivery phase, only the server has access to the database. Every user demands one of the $N$ files. The demand vector is denoted by $\mathbf{d} = (d_1, \ldots, d_K)$, where $d_i$ is the index of the file demanded by user $i$. The number of distinct files requested in $\mathbf{d}$ is denoted by $N_e(\mathbf{d})$. During the delivery phase, the server informed of the demand $\mathbf{d}$, transmits a function of $X_1, \ldots, X_N$, over a shared link. Using the cache contents and the transmitted data, each user $i$ needs to reconstruct the requested file $X_{d_{i}}$ even if $\delta$ transmissions are in error.

For the $\delta$-error correcting coded caching problem, a communication rate $R(\delta)$ is \textit{achievable} for demand $\mathbf{d}$ if and only if there exists a transmission of $R(\delta)F$ bits such that every user $i$ is able to recover its desired file $X_{d_{i}}$ even after at most $\delta$ transmissions are in error. Rate  $R^*(\mathbf{d}, \mathcal{M}, \delta)$ is the minimum achievable rate for a given $\mathbf{d}$, $\mathcal{M}$ and $\delta$. The average rate $R^*(\mathcal{M}, \delta)$ is defined as the expected minimum average rate given $\mathcal{M}$ and $\delta$ under uniformly random demand. Thus $ R^*(\mathcal{M}, \delta) = \mathbb{E}_{\mathbf{d}}[R^*(\mathbf{d}, \mathcal{M}, \delta)].$

The average rate depends on the prefetching scheme $\mathcal{M}$. The minimum average rate  $ R^*(\delta)= \min_{\mathcal{M}} R^*(\mathcal{M}, \delta)$  is the minimum rate of the delivery scheme over all possible $\mathcal{M}$. The rate-memory trade-off for average rate is finding the minimum average rate $R^*(\delta)$ for different memory constraints $M$. Another quantity of interest is the peak rate, denoted by $R^*_{\text{worst}}(\mathcal{M}, \delta)$, which is defined as
$R^*_{\text{worst}}(\mathcal{M}, \delta) = \max_{\mathbf{d}} R^*(\mathbf{d}, \mathcal{M}, \delta).$ 
The minimum peak rate is defined as
$ R^*_{\text{worst}}(\delta)= \min_{\mathcal{M}} R^*_{\text{worst}}(\mathcal{M}, \delta).$

\subsection{Ali-Niesen Decentralized Scheme}
In this subsection, we briefly present the decentralized coded caching scheme in \cite{MaN2}. We denote the decentralized prefetching scheme as $\mathcal{M}_{\text{D}}.$ During the placement phase, each user independently caches a subset of $\frac{MF}{N}$ bits of each file, chosen uniformly at random. Hence, each bit of a file is cached by a specific user with a probability $M/N$. The actions of the placement procedure effectively partition each file $X_i$ into $2^K$ subfiles of the form $X_{i,\mathcal{S}},$ where for $\mathcal{S} \subseteq [K]$, denotes the bits of $X_i$ that are stored exclusively in the cache memories of users in $\mathcal{S}$. Moreover, for large file size $F$, by the law of large numbers
$|X_{i,\mathcal{S}}|\approx (M/N)^{|\mathcal{S}|}(1-M/N)^{K-|\mathcal{S}|}F.$ Two delivery procedures are proposed in \cite{MaN2}, of which the one which is used for the $N \geq K$ regime is as follows. For $\mathcal{S} \subset [K]$ and $|\mathcal{S}|=s$, the server transmits $\oplus_{k \in \mathcal{S}} V_{k,\mathcal{S}\setminus \{k\}} $ for $s=K,K-1, \ldots, 1$. Here $V_{k,\mathcal{S}}$ denotes the bits of file $X_{d_k}$ requested by user $k$ cached exclusively at users in $\mathcal{S}.$
This delivery scheme achieves the rate  
$$R(\mathcal{M}_{\text{D}},0) = (1-M/N)\frac{N}{M}(1-(1-M/N)^K), $$
for $N \geq K$ or $M \geq 1$ regime. 

\subsection{Equivalent Index Coding Problems of a Coded Caching Problem}
For a fixed prefetching $\mathcal{M}$ and for a fixed demand $\mathbf{d}$, the delivery phase of a coded caching problem is an index coding problem \cite{MaN}. In fact, for fixed prefetching, a coded caching scheme consists of $N^K$  parallel index coding problems one for each of the $N^K$ possible user demands. Thus finding the minimum achievable rate for a given demand $\mathbf{d}$ is equivalent to finding the min-rank of the equivalent index coding problem induced by the demand $\mathbf{d}$.

Consider an index coding problem with $\alpha(\mathcal{I}) = \kappa(\mathcal{I})$. For this problem, the optimal construction of error correcting index code is by concatenation of a smallest length index code with an optimal error correcting code. For problems with $\alpha(\mathcal{I}) \neq \kappa(\mathcal{I})$, the optimal construction of error correcting index codes is unknown. Thus concatenation scheme for the construction of optimal error correcting index code may not be optimal in general, which is summarized as follows.
\begin{itemize}
	\item For index coding problems with $\alpha(\mathcal{I}) \neq \kappa(\mathcal{I})$, the concatenation scheme is not proven to be optimal even if the minimum length index code is known.
	\item For index coding problems with $\alpha(\mathcal{I}) = \kappa(\mathcal{I})$, if an optimal index code is not known, then concatenating a non-optimal index code with an optimal error correcting code is not optimal.
	\item For index coding problems with $\alpha(\mathcal{I}) = \kappa(\mathcal{I})$ and if an optimal index code is known, then concatenating it with an optimal error correcting code is optimal.
\end{itemize}

Hence, if for some problems, we have an optimal index code and if for such problems, $\alpha(\mathcal{I}) = \kappa(\mathcal{I})$, then the concatenation scheme is optimal.
In our work, we consider all the index coding problems corresponding to the worst case demands of Ali-Niesen decentralized scheme for $N \geq K$.
For all these index coding problems, we find closed form expression for $\alpha(\mathcal{I})$. The number of bits transmitted in Ali-Niesen delivery scheme turns out to be same as the number of bits corresponding to $\alpha(\mathcal{I})$. Since $\alpha(\mathcal{I}) \leq \kappa(\mathcal{I})$ in general \cite{KTR}, we get  $\alpha(\mathcal{I}) = \kappa(\mathcal{I})$ for all the corresponding index coding problems.
Hence, for all these problems, concatenation of Ali-Niesen delivery scheme with an optimal error correcting code gives an optimal error correcting delivery scheme.

 The length of an optimal linear $(\delta, \mathcal{I})$-ECIC over $\mathbb{F}_q$ satisfies \eqref{eq:bds}.  Whatever be the combinations of index codes and error correcting codes being tried, the length of ECIC should be greater than or equal to the $\alpha$-bound. The constructions of error correcting delivery schemes  used in this paper are in such a way that their lengths meet the $\alpha$-bound with equality and thus are optimal.

It is shown in the Example 4.8 of \cite{DSC} that the inequality can be strict in general. In particular, it follows that mere application of an optimal length error-correcting code on top of an optimal index code may fail to provide us with an optimal linear ECIC. This example is reproduced here for convenience. 
\begin{exmp}
	Let field size $q=2$ and number of messages, $n=5$. Let $\delta =2$ errors need to be corrected and let the demand of $i$th receiver be $x_i$, for $1 \leq i \leq 5$. Let the side information sets be given as $\mathcal{X}_1=\{2,5\}$,  $\mathcal{X}_2=\{1,3\}$, $\mathcal{X}_3=\{2,4\}$, $\mathcal{X}_4=\{3,5\}$ and $\mathcal{X}_5=\{1,4\}$. The set $\mathcal{J}(\mathcal{I})$ for this problem is given by
	\begin{align*}
	\mathcal{J}(\mathcal{I}) = \{& \{1\}, \{1,3\},\{1,4\}, \{1,3,4\}, \\
	& \{2\}, \{2,4\},\{2,5\}, \{2,4,5\}, \\ 
	&\{3\}, \{1,3\},\{3,5\}, \{1,3,5\}, \\
	& \{4\}, \{1,4\},\{2,4\}, \{1,2,4\}, \\
	& \{5\}, \{2,5\},\{3,5\}, \{2,3,5\} \}.
	\end{align*}
	For this problem it can be calculated that $\alpha(\mathcal{I})=2$. Also, for this problem, min-rank, $\kappa(\mathcal{I})=3$. From code tables in \cite{Gra}, we have $N_2[2,5]=8$ and $N_2[3,5]=10.$ Hence,  $8 \leq N_2[\mathcal{I}, \delta] \leq 10$. Using a computer search, the authors of [10] have found that the optimal length $\mathcal{N}_2[\mathcal{I},2]=9$. Here the optimal length of the ECIC lies strictly between the $\alpha$-bound and the $\kappa$-bound.
\end{exmp}

Consider the decentralized prefetching scheme $\mathcal{M}_{\text{D}}$. The index coding problem induced by the demand $\mathbf{d}$  for the decentralized prefetching is denoted by $\mathcal{I}(\mathcal{M}_{\text{D}}, \mathbf{d}).$  The corresponding generalized independence number and min-rank are represented as $\alpha(\mathcal{M}_{\text{D}}, \mathbf{d})$ and $\kappa(\mathcal{M}_{\text{D}}, \mathbf{d})$ respectively. Also, the demand vector $\mathbf{d}_{\text{worst}}$ corresponds to the case when all the demanded files are distinct.
\label{sec:index_caching}


\section{Optimality of the Ali-Niesen Decentralized Scheme}
\label{sec:optimality}
In this section we prove that the decentralized scheme in \cite{MaN2} is optimal for $N \geq K$ using results from index coding. Moreover, the results presented in this section are used to construct optimal error correcting delivery scheme for coded caching problem with decentralized  prefetching in Section \ref{sec:error_cor}. The theorem below gives a lower bound for $\alpha({\mathcal{M}_{D}}, \mathbf{d}_{\text{worst}})$ which is used to prove the optimality of the Ali-Niesen decentralized scheme.  

\begin{thm}
	For the index coding problem $\mathcal{I}(\mathcal{M}_{D}, \mathbf{d}_\text{worst})$ corresponding to a coded caching problem with Ali-Niesen decentralized prefetching for $N \geq K$ and the worst case demand $\mathbf{d}_\text{worst}$, 
	\begin{align*}
	\alpha({\mathcal{M}_{D}, \mathbf{d}_\text{worst}}) &=\kappa({\mathcal{M}_{D}, \mathbf{d}_\text{worst}}) \\
	&=(1-M/N)\frac{N}{M}\big(1-(1-M/N)^K\big)F.
	\end{align*}
	Thus, the decentralized scheme in \cite{MaN2} is optimal for $N \geq K$.
	\label{Thm:dec_alph_kap}
\end{thm}

\begin{IEEEproof}
	Worst case demand scenario is considered here. Without loss of generality we can assume that the demand vector is $\mathbf{d}=(1,2, \ldots, K).$ The corresponding index coding problem $\mathcal{I}(\mathcal{M}_{D}, \mathbf{d}_\text{worst})$ can be viewed to be consisting of $NF$ messages each of one bit and $KF$ receivers demanding $KF$ bits corresponding to the first $K$ files. We construct a set $B(\mathcal{I})$, whose elements are messages of the index coding problem such that the set of indices of the messages in $B(\mathcal{I})$ forms a generalized independent set. The set $B(\mathcal{I})$ is constructed as 
	$$  B(\mathcal{I})= \bigcup_{i \in [K]} \{X_{i, \mathcal{S}}: 1,2, \ldots, i \notin \mathcal{S}\}.$$
	Let $H(\mathcal{I})$ be the set of indices of the messages in $B(\mathcal{I})$. The claim is that  $H(\mathcal{I})$ is a generalized independent set. Each message in $B(\mathcal{I})$ is demanded by one receiver. Hence all the subsets of $H(\mathcal{I})$ of size one are present in $\mathcal{J}(\mathcal{I})$.
	Consider any set $C = \{X_{i_1, \mathcal{S}_1}, X_{i_2, \mathcal{S}_2}, \ldots, X_{i_k, \mathcal{S}_k}\} \subseteq B(\mathcal{I})$ where $i_1 \leq i_2 \leq \ldots \leq i_k$. Consider the message $X_{i_1, \mathcal{S}_1}$. The receiver demanding this message does not have any other message in $C$ as side information. Thus indices of messages in $C$ lie in $\mathcal{J}(\mathcal{I})$. Thus any subset of $H(\mathcal{I})$ lies in $\mathcal{J}(\mathcal{I})$. 	Since $H(\mathcal{I})$ is a generalized independent set, we have, $\alpha({\mathcal{M}_{D}, \mathbf{d}_\text{worst}}) \geq |H(\mathcal{I})| $. Note that $|H(\mathcal{I})|=|B(\mathcal{I})|$.
There are ${K-n \choose i}$ subfiles of the form $X_{n, \mathcal{S}}$ such that $|\mathcal{S}|=i$ in $ B(\mathcal{I})$. Hence, the number of bits of file $X_n$ of the form $X_{n,S}$ such that $|S|=i$ in $B(\mathcal{I})$ is 
	${K-n \choose i} (M/N)^i(1-M/N)^{K-i}F. $
	Thus,
	\begin{align*}
	|B(\mathcal{I})|&=\sum_{n=1}^{K} \sum_{i=0}^{K-n} {K-n \choose i} (M/N)^i(1-M/N)^{K-i}N \\
	&=  \sum_{n=1}^{K} \sum_{i=0}^{K-n} {K-n \choose i} (M/N)^i(1-M/N)^{K-n-i} \\ & ~~~~~~~~~~~~~~~~~~~~~~~~~~~~~~~~~~~~~~~~(1-M/N)^nF \\
	&= \sum_{n=1}^{K} (1-M/N)^nF.
	\end{align*}
	The last equality follows from the binomial expansion $ (x+y)^n= \sum_{k=0}^{n} {n \choose k} x^n y^{n-k}. $ The expression $\sum_{n=1}^{K} (1-M/N)^n$ is sum of a geometrical progression.
	Thus,
	\begin{align*}
	|B(\mathcal{I})| &= \frac{(1-M/N)(1-(1-M/N)^K)}{(1-(1-M/N))}F \\
	&= (1-M/N).\frac{N}{M}(1-(1-M/N)^K)F.     
	\end{align*}
	Thus, $\alpha({\mathcal{M}_{D}, \mathbf{d}_\text{worst}}) \geq (1-M/N).\frac{N}{M}(1-(1-M/N)^K)F.$ Also, from the achievable scheme in \cite{MaN2}, it follows that, $\kappa({\mathcal{M}_{D}, \mathbf{d}_\text{worst}}) \leq (1-M/N).\frac{N}{M}(1-(1-M/N)^K)F.$
The rate achieved by the scheme in \cite{MaN2} thus meets the lower bound, which proves its optimality.	Hence the statement of the theorem follows.
\end{IEEEproof}

Two examples are given below to illustrate the construction of generalized independent set for the index coding problems corresponding to coded caching problem with decentralized placement.

\begin{exmp}
	\label{ex:NK2}
	Consider a coded caching problem with $N=2$ files and $K=2$ users each with a cache of size $M \in [0,2]$. In the placement phase, each user caches a subset of $MF/2$ bits of each file independently at random. Thus both the files $X_1$ and $X_2$ are effectively partitioned into four subfiles:
\begin{align*}
X_1 &=(X_{1,\phi}, X_{1,\{ 1\}}, X_{1,\{2\}}, X_{1,\{ 1,2\}}) \text{ and } \\
X_2 &=(X_{2,\phi}, X_{2,\{ 1\}}, X_{2,\{2\}}, X_{2,\{ 1,2\}}).
\end{align*}
	For large enough file size $F$, we have with high probability for $i \in [2]$,
\begin{align*}
|X_{i,\phi}|  &= (1-M/2)^2F, \\
	|X_{i,\{ 1\}}| &= |X_{i,\{ 2\}}| = (M/2)(1-M/2)F \text{ and } \\
	 |X_{i,\{ 1,2\}}|&=(M/2)^2F. 
\end{align*}
	Let the demand vector be $\mathbf{d}=(1,2).$ Let the corresponding index coding problem be $\mathcal{I}(\mathcal{M}_{D}, \mathbf{d}).$ A generalized independent set can be constructed for this problem following the procedure in the proof of Theorem \ref{Thm:dec_alph_kap} as
	$$ B(\mathcal{I})= \{X_{1,\phi}, X_{1,\{2\}}, X_{2,\phi}\}. $$
	From this, $|B(\mathcal{I})|=2(1-M/2)^2F+ (M/2)(1-M/2)F.$
	Thus $\alpha(\mathcal{M}_{D}, \mathbf{d}) \geq 2(1-M/2)^2F+ (M/2)(1-M/2)F. $
	From \cite{MaN2}, we have the number of bits transmitted is exactly $2(1-M/2)^2F+ (M/2)(1-M/2)F.$ Hence $\kappa(\mathcal{M}_{D}, \mathbf{d}) \leq 2(1-M/2)^2F+ (M/2)(1-M/2)F.$ Hence $\alpha({\mathcal{M}_{D}, \mathbf{d}}) =\kappa({\mathcal{M}_{D}, \mathbf{d}}).$ The set $B(\mathcal{I})$ can be constructed for the two possible combinations of distinct demands and are shown in Table \ref{Tab:Ex_alpha1}. For both the cases, the cardinality of $B(\mathcal{I})$ turns out to be same, which is equal to $\kappa({\mathcal{M}_{D}, \mathbf{d}}).$
	
	\begin{table}
		\center
		\begin{tabular} {|c|c| }
			\hline
			Demand $\mathbf{d}$ & $B(\mathcal{I})$ \\ 
			\hline
			$(1,2)$ & $\{X_{1,\phi}, X_{1,\{2\}},  X_{2,\phi}\}$  \\
			\hline
			$(2,1)$ & $\{X_{2,\phi},X_{2,\{2\}},  X_{1,\phi}\}$ \\
			\hline
		\end{tabular}
		\newline
		\caption{Generalized independent sets of $\mathcal{I}(\mathcal{M}_{D}, {\mathbf{d}_{\text{worst}}})$ for different demands for Example \ref{ex:NK2}}
		\label{Tab:Ex_alpha1}
	\end{table}		  
\end{exmp}
\begin{exmp}
	\label{ex:N4K3}
	Consider a coded caching problem with $N=4$ files and $K=3$ users. Let $M=1$. In the placement phase, each user caches a subset of $MF/N=F/4$ bits of each file independently at random. Thus each file is effectively partitioned as follows:
	\begin{align*}
	X_i =( X_{i, \phi}, X_{i, \{1\}}, X_{i, \{2\}}, X_{i, \{3\}}, X_{i, \{1,2\}}
	, X&_{i,\{1,3\}}, X_{i, \{2,3\}},\\ & X_{i, \{1,2,3\}}),
	\end{align*}
	for $i \in [4]$.
	For sufficiently large $F$, the number of bits corresponding to each of these subfiles is given as:
\begin{align*}
 |X_{i,\phi}|  &= \frac{27}{64}F, \\
 |X_{i,\{1\}}| &= |X_{i,\{2\}}|= |X_{i,\{3\}}|= \frac{9}{64}F ,\\
  |X_{i,\{1,2\}}|  &= |X_{i,\{1,3\}}|= |X_{i,\{2,3\}}|= \frac{3}{64}F \text{ and } \\
	 |X_{i,\{1,2,3\}}|&= \frac{1}{64}F.  
\end{align*}
	Let the demand vector be $\mathbf{d}=(1,2,3).$ A generalized independent set of the corresponding index coding problem can be constructed for following the procedure in the proof of Theorem \ref{Thm:dec_alph_kap} as
	$$ B(\mathcal{I})= \{X_{1,\phi}, X_{1,\{2\}},  X_{1,\{3\}},  X_{1,\{2,3\}},  X_{2,\phi},  X_{2,\{3\}}, X_{3,\phi}\}. $$
	Hence, $|B(\mathcal{I})|=\frac{111}{64}F$.
	Thus $\alpha(\mathcal{M}_{D}, \mathbf{d}) \geq \frac{111}{64}F. $ 
	From \cite{MaN2}, we have the number of bits transmitted is $\frac{111}{64}F.$ Hence $\kappa(\mathcal{M}_{D}, \mathbf{d}) \leq \frac{111}{64}F.$ Hence $\alpha({\mathcal{M}_{D}, \mathbf{d}}) =\kappa({\mathcal{M}_{D}, \mathbf{d}}).$   
	 The set $B(\mathcal{I})$ can be constructed for all the possible combinations of distinct demands and are shown in Table \ref{Tab:Ex_alpha2}. For all the cases, the cardinality of $B(\mathcal{I})$ turns out to be same, which is equal to $\kappa({\mathcal{M}_{D}, \mathbf{d}}).$
	\begin{table}
		\center
		\begin{tabular} {|c|c| }
			\hline
			Demand $\mathbf{d}$ & $B(\mathcal{I})$ \\ 
			\hline
			$(1,2,3)$ & $\{X_{1,\phi}, X_{1,\{2\}}, X_{1,\{3\}}, X_{1,\{2,3\}},  X_{2,\phi}, X_{2,\{3\}},X_{3,\phi}\}$  \\
			\hline
			$(1,3,2)$ & $\{X_{1,\phi}, X_{1,\{2\}}, X_{1,\{3\}}, X_{1,\{2,3\}},  X_{3,\phi}, X_{3,\{3\}},X_{2,\phi}\}$  \\
			\hline
			$(2,1,3)$ & $\{X_{2,\phi}, X_{2,\{2\}}, X_{2,\{3\}}, X_{2,\{2,3\}},  X_{1,\phi}, X_{1,\{3\}},X_{3,\phi}\}$  \\
			\hline
			$(2,3,1)$ & $\{X_{2,\phi}, X_{2,\{2\}}, X_{2,\{3\}}, X_{2,\{2,3\}},  X_{3,\phi}, X_{3,\{3\}},X_{1,\phi}\}$  \\
			\hline
			$(3,1,2)$ & $\{X_{3,\phi}, X_{3,\{2\}}, X_{3,\{3\}}, X_{3,\{2,3\}},  X_{1,\phi}, X_{1,\{3\}},X_{2,\phi}\}$  \\
			\hline
			$(3,2,1)$ & $\{X_{3,\phi}, X_{3,\{2\}}, X_{3,\{3\}}, X_{3,\{2,3\}},  X_{2,\phi}, X_{2,\{3\}},X_{1,\phi}\}$  \\
			\hline
			$(1,2,4)$ & $\{X_{1,\phi}, X_{1,\{2\}}, X_{1,\{3\}}, X_{1,\{2,3\}},  X_{2,\phi}, X_{2,\{3\}},X_{4,\phi}\}$  \\
			\hline
			$(1,4,2)$ & $\{X_{1,\phi}, X_{1,\{2\}}, X_{1,\{3\}}, X_{1,\{2,3\}},  X_{4,\phi}, X_{4,\{3\}},X_{2,\phi}\}$  \\
			\hline
			$(2,1,4)$ & $\{X_{2,\phi}, X_{2,\{2\}}, X_{2,\{3\}}, X_{2,\{2,3\}},  X_{1,\phi}, X_{1,\{3\}},X_{4,\phi}\}$  \\
			\hline
			$(2,4,1)$ & $\{X_{2,\phi}, X_{2,\{2\}}, X_{2,\{3\}}, X_{2,\{2,3\}},  X_{4,\phi}, X_{4,\{3\}},X_{1,\phi}\}$  \\
			\hline
			$(4,1,2)$ & $\{X_{4,\phi}, X_{4,\{2\}}, X_{4,\{3\}}, X_{4,\{2,3\}},  X_{1,\phi}, X_{1,\{3\}},X_{2,\phi}\}$  \\
			\hline
			$(4,2,1)$ & $\{X_{4,\phi}, X_{4,\{2\}}, X_{4,\{3\}}, X_{4,\{2,3\}},  X_{2,\phi}, X_{2,\{3\}},X_{1,\phi}\}$  \\
			\hline
		$(1,4,3)$ & $\{X_{1,\phi}, X_{1,\{2\}}, X_{1,\{3\}}, X_{1,\{2,3\}},  X_{4,\phi}, X_{4,\{3\}},X_{3,\phi}\}$  \\
		\hline
		$(1,3,4)$ & $\{X_{1,\phi}, X_{1,\{2\}}, X_{1,\{3\}}, X_{1,\{2,3\}},  X_{3,\phi}, X_{3,\{3\}},X_{4,\phi}\}$  \\
		\hline
		$(4,1,3)$ & $\{X_{4,\phi}, X_{4,\{2\}}, X_{4,\{3\}}, X_{4,\{2,3\}},  X_{1,\phi}, X_{1,\{3\}},X_{3,\phi}\}$  \\
		\hline
		$(4,3,1)$ & $\{X_{4,\phi}, X_{4,\{2\}}, X_{4,\{3\}}, X_{4,\{2,3\}},  X_{3,\phi}, X_{3,\{3\}},X_{1,\phi}\}$  \\
		\hline
		$(3,1,4)$ & $\{X_{3,\phi}, X_{3,\{2\}}, X_{3,\{3\}}, X_{3,\{2,3\}},  X_{1,\phi}, X_{1,\{3\}},X_{4,\phi}\}$  \\
		\hline
		$(3,4,1)$ & $\{X_{3,\phi}, X_{3,\{2\}}, X_{3,\{3\}}, X_{3,\{2,3\}},  X_{4,\phi}, X_{4,\{3\}},X_{1,\phi}\}$  \\
		\hline
		$(4,2,3)$ & $\{X_{4,\phi}, X_{4,\{2\}}, X_{4,\{3\}}, X_{4,\{2,3\}},  X_{2,\phi}, X_{2,\{3\}},X_{3,\phi}\}$  \\
		\hline
		$(4,3,2)$ & $\{X_{4,\phi}, X_{4,\{2\}}, X_{4,\{3\}}, X_{4,\{2,3\}},  X_{3,\phi}, X_{3,\{3\}},X_{2,\phi}\}$  \\
		\hline
		$(2,4,3)$ & $\{X_{2,\phi}, X_{2,\{2\}}, X_{2,\{3\}}, X_{2,\{2,3\}},  X_{4,\phi}, X_{4,\{3\}},X_{3,\phi}\}$  \\
		\hline
		$(2,3,4)$ & $\{X_{2,\phi}, X_{2,\{2\}}, X_{2,\{3\}}, X_{2,\{2,3\}},  X_{3,\phi}, X_{3,\{3\}},X_{4,\phi}\}$  \\
		\hline
		$(3,4,2)$ & $\{X_{3,\phi}, X_{3,\{2\}}, X_{3,\{3\}}, X_{3,\{2,3\}},  X_{4,\phi}, X_{4,\{3\}},X_{2,\phi}\}$  \\
		\hline
		$(3,2,4)$ & $\{X_{3,\phi}, X_{3,\{2\}}, X_{3,\{3\}}, X_{3,\{2,3\}},  X_{2,\phi}, X_{2,\{3\}},X_{4,\phi}\}$  \\
		\hline
		\end{tabular}
		\newline
		\caption{Generalized independent sets of $\mathcal{I}(\mathcal{M}_{D}, {\mathbf{d}_{\text{worst}}})$ for different demands for Example \ref{ex:N4K3}}
		\label{Tab:Ex_alpha2}
	\end{table}		  
\end{exmp}

\section{Optimal Error Correcting Delivery Scheme for Ali-Niesen Decentralized Prefetching}
\label{sec:error_cor}
For the worst case demand, we have proved in Theorem \ref{Thm:dec_alph_kap} that $\alpha(\mathcal{M}_{\text{D}},\mathbf{d}_{\text{worst}}) = \kappa(\mathcal{M}_{\text{D}},\mathbf{d}_{\text{worst}})$. Hence for this case, the optimal linear error correcting delivery scheme can be constructed by concatenating the delivery scheme in \cite{MaN2} with an optimal error correcting code which corrects the required number of errors. Based on this we give an expression for the worst case rate for decentralized prefetching in the theorem below.

\begin{thm}
	For a coded caching problem with Ali-Niesen decentralized prefetching for $N \geq K$,
	$$ R^*_\text{worst}(\mathcal{M}_{D}, \delta) = \frac{N_q[\kappa({\mathcal{M}_{D}, \mathbf{d}_{\text{worst}}}), 2\delta+1]}{F},$$
	where $ \kappa({\mathcal{M}_{D}, \mathbf{d}_{\text{worst}}}) = (1-M/N)\frac{N}{M}\big(1-(1-M/N)^K\big)F.$ 
\end{thm}
\begin{IEEEproof}
	From Theorem \ref{Thm:dec_alph_kap}, we have for any index coding problem corresponding to coded caching with decentralized placement for $N \geq K$, $\alpha({\mathcal{M}_{D}, \mathbf{d}_{\text{worst}}}) = \kappa({\mathcal{M}_{D}, \mathbf{d}_{\text{worst}}}).$ Hence by \eqref{eq:bds}, the $\alpha$ and $\kappa$ bounds become equal for all these index coding problems. Thus, the optimal number of transmissions required for $\delta$ error corrections in those index coding problems is 
	$N_q[\kappa({\mathcal{M}_{D}, \mathbf{d}_{\text{worst}}}), 2\delta+1].$ Hence the statement of theorem follows.
\end{IEEEproof} 

Since the $\alpha$ and $\kappa$ bounds meet for all the index coding problems corresponding to the worst case demands of coded caching problems with decentralized placement for $N=K$, the optimal coded caching delivery scheme would be the concatenation of the delivery scheme proposed in \cite{MaN2} with an optimal classical error correcting code which corrects $\delta$ errors. Decoding is done by syndrome decoding for error correcting index codes proposed in \cite{DSC}.

\begin{exmp}
	Consider the decentralized coded caching problem with $N=K=2$ considered in Example \ref{ex:NK2}. Consider the case when $M=1$. We use the decentralized delivery scheme \cite{MaN2} for this example. In this case, all the subfiles of the form $X_{i, \mathcal{S}}$ have the same number of bits for large $F$. Hence for simplicity, these subfiles can be considered as the messages of the corresponding index coding problem. The transmissions given in \cite{MaN2} are
	$$ X_{1,\{2\}} \oplus  X_{2,\{1\}},~~~ X_{1,\phi}~~ \text{ and } ~~ X_{2,\phi}.$$
	If $\delta=1$ error need to be corrected, the optimal scheme is to concatenate these transmissions with a classical error correcting code of optimal length. From \cite{Gra}, we have $N_2[3,3]=6$. One such code is given by the following generator matrix
	$$
	\bf{G}=
	\begin{bmatrix}
	1 & 0 & 0 & 1 & 1 & 0 \\
	0 & 1 & 0 & 1 & 0 & 1 \\
	0 & 0 & 1 & 0 & 1 & 1
	\end{bmatrix}.
	$$
	Concatenating this $[6,3,3]_2$ code with the decentralized transmissions give rise to six transmissions given by:
\begin{align*}
T_1 &: X_{1,\{2\}} \oplus  X_{2,\{1\}},  \\ 
T_2 &: X_{1,\phi}, \\  
T_3 &: X_{2,\phi}, \\
T_4 &:  X_{1,\{2\}} \oplus  X_{2,\{1\}} \oplus X_{1,\phi}, \\
 T_5 &: X_{1,\{2\}} \oplus  X_{2,\{1\}} \oplus X_{2,\phi} \text{ and } \\
	 T_6 &: X_{1,\phi} \oplus X_{2,\phi}.   
\end{align*}

	Decoding is done by syndrome decoding for error correcting index codes proposed in \cite{DSC}. The number of bits involved in each transmission is $F/4$. Hence, the rate of transmission is $3/2$. For zero error correcting scenario the rate corresponding to $M=1$ was $3/4$.
\end{exmp}

\begin{exmp}
	Consider the decentralized coded caching problem with $N=4$ and $K=3$ considered in Example \ref{ex:N4K3}. Consider the case when $M=1$. We use the decentralized delivery scheme \cite{MaN2} for this example. For simplicity we consider that for the corresponding index coding problem, each index coding message is of $F/64$ bits. The transmissions given in \cite{MaN2} are:
	 $X_{1,\{2,3\}} \oplus X_{2,\{1,3\}} \oplus X_{3,\{1,2\}}, X_{1,\{2\}}\oplus X_{2,\{1\}}, X_{2,\{3\}} \oplus X_{3,\{2\}}, X_{1,\{3\}}\oplus X_{3,\{1\}}, X_{1,\phi}, X_{2,\phi} \text{ and } X_{3,\phi}.$ 
	
	The total number of bits transmitted is $\frac{111F}{64}.$ If each index coding message is considered  consisting of $F/64$ bits, the min-rank $\kappa({\mathcal{M}_{\text{D}}, \mathbf{d}_{\text{worst}}})=111.$ 
	If $\delta=1$ error need to be corrected, the optimal scheme is to concatenate these transmissions with a classical error correcting code of optimal length. From \cite{Gra}, we have $N_2[111,3]=118$. 
	Hence, concatenating $[118,111,3]_2$ code with the decentralized transmissions give rise to optimal error correcting delivery scheme.
	Decoding is done by syndrome decoding for error correcting index codes proposed in \cite{DSC}. 
\end{exmp}

\section{Conclusion}
We considered the decentralized coded caching problem in \cite{MaN2} and proved that for $N \geq K$, the delivery scheme in \cite{MaN2} is optimal using the results from index coding. Also, since the $\alpha$ and $\kappa$ bounds meet for the corresponding index coding problems, the concatenation of decentralized delivery scheme with an optimal classical error correcting code which corrects the required number of errors is optimal. A good direction of future work would be to find the optimality of other coded caching schemes by finding the generalized independence numbers for the corresponding index coding problems. For the case of non-optimal schemes, the gap from optimality can be measured if the generalized independence number is found out.


\section*{Acknowledgment}
This work was supported partly by the Science and Engineering Research Board (SERB) of Department of Science and Technology (DST), Government of India, through J.C. Bose National Fellowship to B. Sundar Rajan.

\end{document}